\documentstyle[epsfig]{article}
\def\three_j(#1,#2,#3,#4,#5,#6){\pmatrix{#1 & #2 & #3\cr
                                         #4 & #5 & #6\cr}}

\def\qqq{\end{document}}
\def\pmb#1{\setbox0=\hbox{$#1$}%
\kern-.025em\copy0\kern-\wd0
\kern.05em\copy0\kern-\wd0
\kern-.025em\raise.0433em\box0 }

\def\xara(#1,#2,#3,#4){\left(\matrix{#1 & #2\cr #3 & #4\cr}\right)}

\def\six_j(#1,#2,#3,#4,#5,#6){\left\{\matrix{#1 & #2 & #3\cr
                                         #4 & #5 & #6\cr}\right\}}
\def\nine_j(#1,#2,#3,#4,#5,#6,#7,#8,#9){\left\{\matrix{#1 & #2 & #3\cr
                                        #4 & #5 & #6\cr
                                         #7 & #8 & #9\cr}\right\}}

\def\Ener(#1,#2){ \sqrt{{#1}^2+{#2}^2} }

\def\overlay#1#2{\setbox0=\hbox{$#1$}\setbox1=\hbox to \wd0{\hss$#2$\hss}#1%
\hskip -1\wd0\copy1}

\def\bold#1{\setbox0=\hbox{$#1$}%
      \kern-.025em\copy0\kern-\wd0
      \kern.05em\copy0\kern-\wd0
      \kern-.025em\raise.0433em\box0 }

\def\S11{S_{11}(1535)}
\def\E0+{E_{0^+}}

\def\footnoterule{\kern-3pt \hrule width \hsize \kern2.6pt}

\newcommand{\be}{\begin{equation}}
\newcommand{\ee}{\end{equation}}
\newcommand{\ba}{\begin{eqnarray}}
\newcommand{\ea}{\end{eqnarray}}

\newcommand{\bd}[1]{ \mbox{\boldmath $#1$}  }

\begin{document}
\begin{center}
{\Large \bf 
%
Relativistic current densities for bound spin--orbit partners 
and the longitudinal--transverse response in (e,e$'$p) processes
%
}\\[1cm]
J.A. Caballero$^{1,2}$, 
T.W. Donnelly$^{3}$,
E. Moya de Guerra$^{2}$ and 
J.M. Ud\'\i as$^{4}$
\\[0.7cm]
$^{1}${\sl 
Departamento de F\'\i sica At\'omica, Molecular y Nuclear \\ 
Universidad de Sevilla, Apdo. 1065, E-41080 Sevilla, SPAIN 
}\\[5mm]
$^{2}${\sl 
Instituto de Estructura de la Materia, Consejo Superior de
Investigaciones Cient\'{\i}ficas \\ 
Serrano 123, E-28006 Madrid, SPAIN 
}\\[5mm]
$^{3}${\sl 
Center for Theoretical Physics, Laboratory for Nuclear Science and
Dept. of Physics\\
Massachusetts Institute of Technology,
Cambridge, MA 02139, USA 
}\\[5mm]
$^{4}${\sl 
Departamento de F\'\i sica At\'omica, Molecular y Nuclear \\ 
Universidad Complutense de Madrid, E-28040 Madrid, SPAIN
}
\end{center}

\vspace{0.5cm}


\begin{abstract}


Scalar, baryon and vector--current densities in coordinate and momentum
space are calculated within a relativistic mean--field model. The role of
the low components of the bound nucleon wave function is investigated in
detail for different spin--orbit partner shells. We show that the
relative importance of the negative--energy projection components can
be  explained from the different quantum numbers
involved in the relativistic wave function. This fact has proved to be
important in the analysis of various electron scattering observables even
for low--medium values of the bound nucleon momentum.

\end{abstract}

\vspace{1.0cm}

\vfill
MIT-CTP\#2764 \hfill August, 1998

\newpage


\section*{1. Introduction}


In a recent publication~\cite{Cab98} we investigated the role of the
negative--energy projections (NEP) of the relativistic bound nucleon wave
function in the longitudinal (L), transverse (T, TT) and
longitudinal--transverse (TL) response functions of (e,e$'$p) processes.
Our work was motivated by the success in the interpretation of
recent $(e,e'p)$  data~\cite{data}
of the relativistic distorted wave impulse approximation 
(RDWIA)~\cite{Udi93,Udi95,Jin92} that uses  relativistic
bound and scattering wave functions (based on the S--V Walecka
model~\cite{Wal74,Ser86} and Dirac phenomenology). 
Data on momentum distributions obtained 
 from exclusive quasielastic experiments over
extended ranges of missing momenta at low missing energies 
have been successfully analyzed in RDWIA
 producing reasonable
values of the spectroscopic factors ($s_\alpha\sim 0.7$ for the
 $s_{1/2}$ shell in $^{208}Pb$) consistent at low and high
momentum~\cite{Udi93,Udi96}. 

The study carried out  in ref.~\cite{Cab98} was made in the
relativistic plane wave impulse approximation (RPWIA) that differs
from RDWIA in that the scattered nucleon is described as a plane
wave. Both RPWIA and RDWIA calculations can also be found
in refs.~\cite{npa15,npa16,npa14}. The present work focusses on the TL
response in RPWIA for the reasons discussed below.

The investigation in ref.~\cite{Cab98} was focused on the relativistic 
plane--wave impulse
approximation (RPWIA) with the purpose of isolating effects due to the
relativistic bound nucleon wave function and current. We discussed the
results for three choices of the nucleon current gauge: Landau or NCCi, Coulomb
or CCi(0) and Weyl or CCi(3). We also considered two current operators 
that are called
i=1,2 following de Forest's notation~\cite{deF83}. It was found that the
role of the NEP of the relativistic bound nucleon wave function in RPWIA
is to cause a reduction of the differential
cross section at low $p$ ($p\leq 300$ MeV) and an increase 
at high $p$ ($p\geq 300$ MeV).
In RPWIA~\cite{Cab98}, the
reduction at low $p$ is generally small and depends little on the current
operator and gauge choice with one proviso: the Weyl gauge choice may
produce up to $20\%$ deviations from Landau and Coulomb gauges at low $p$.
However this deviation is also present when the NEP of the relativistic
bound nucleon wave functions is neglected. It is known to be due to the
difference between $\sigma_{CC(0)}$ and $\sigma_{CC(3)}$ defined in
ref.~\cite{deF83}, therefore appearing even when non--relativistic (n.r.) 
bound nucleon wave functions are used~\cite{Naus}.
At larger $p$ values ($p\geq 300$ MeV) the size of the NEP contributions
depends largely on the choice of the current operator and gauge. The NEP
contributions are minimized for the Landau and Coulomb gauge and for
i=2 current operator, while the Weyl gauge and the i=1 current
operator tend to maximize the effect of the NEP contributions, producing
large differences in some of the response functions. This is particularly
the case for the TL response function ($R^{TL}$) where we found 
variations of up to a factor of three.

At present a certain degree of controversy surrounds the TL responses obtained
from exclusive quasielastic electron scattering experiments for the least
bound orbitals in several nuclei ($^{12}$C, $^{16}$O, $^{208}$Pb.....):
in some cases~\cite{TL1} large deviations from standard DWIA calculations
appear while in others~\cite{TL2} the data are close to the DWIA
calculations. 

Since
our study in ref.~\cite{Cab98} was focused on the $p_{1/2}$ shell in
$^{16}$O, it is interesting to know to what extent the above mentioned
discrepancy by a factor three remains when knockout from other orbitals 
is considered. Hence in the present work we extend our study to the 
$1p_{3/2}$ orbital (in $^{12}$C and $^{16}$O), as well as to the $1d_{3/2}$, 
$1d_{5/2}$ orbitals in $^{40}$Ca.
 The aim here is to see whether the large
effect on the $R^{TL}$ response found in ref.~\cite{Cab98} is a general feature
present in other shells, or is particular to the $1p_{1/2}$ shell
considered in our recent studies.

With this goal in mind we restrict ourselves to the 
RPWIA, as in our previous work. Of course, before detailed comparisons
can be made with experimental data it is necessary to incorporate the
effects of final--state interactions (FSI) in a full RDWIA approach. 
Although FSI are needed for a quantitative comparison with
experimental data~\cite{Udi93,Udi95,Udi96}, we expect that, at least
for the largest contributions to the cross section (i.e., excluding
from this statement some of the smaller responses, especially some of
those that involve polarization degrees of freedom), the effects of FSI 
to be basically the same
for one--nucleon knockout from different bound orbitals, provided the
outgoing kinetic energy is the same. At the maximum of the
TL response that forms the focus of the present work one finds
 that for instance for the $p_{1/2}$ and $p_{3/2}$ shells in $^{16}O$ the
FSI effects differ at most by a 4\%~\cite{progress}.
We therefore expect that the
conclusions of our study here can be extrapolated to the more realistic
situation when RDWIA calculations are performed. On the other hand the present
study is necessary to identify the dependence on the NEP of the bound nucleon
wave function for different orbitals.

The organization of this paper is as follows: in Section 2 we
introduce, within the context of the Walecka model, the scalar,
baryonic and vector densities in coordinate and momentum
space. Results for various spin--orbit partners are shown and
discussed. In Section 3 we study the transverse--longitudinal response
function for one--nucleon knockout from a selection of relativistic bound
orbitals. We compare and discuss the results obtained for different
current operators and gauges and the role of the negative energy
projections of the relativistic bound nucleon wave function is studied.
Finally in Section 4 we present a summary and our conclusions.


\section*{2. Bound nucleon wave functions and densities 
of spin--orbit partners}


As is well known, in the Walecka model~\cite{Wal74,Ser86} the Dirac
equations for bound nucleons in a finite nucleus are derived from an
interacting relativistic field theory of mesons and baryons by
approximating the meson field operators by classical fields. 
For spherically symmetric spin and isospin saturated nuclei treated
within the context of the Hartree approximation
only scalar, $S=S(r)$, and vector, $V_\mu=(V(r),{\bf 0})$, potentials are
to be considered and we write the time--independent Dirac equation with
$S-V$ potentials as
\be
\left[\gamma_0\tilde{E}-{\bd\gamma}\cdot{\bd p}-\tilde{M}\right]\psi 
       =0\,\,\, ,
\label{eq1}
\ee
with 
\ba
\tilde{M}&=& M-S \,\,\, ,\label{eq2} \\
\tilde{E} &=& E-V \,\,\, ,\label{eq3}
\ea
where $M$ is the nucleon mass and $E$ is the total energy eigenvalue
(that is, $e\equiv M-E>0$ for bound states). 
Here and in what follows the conventions of
Bjorken and Drell~\cite{BD64} are followed. From eq.~(\ref{eq1}) two
Schr\"odinger--like equations can be derived for the upper and lower bispinor
components of $\psi$:
\be
\psi^\kappa =
\left(\begin{array}{@{\hspace{0pt}}c@{\hspace{0pt}}}
                \psi^\kappa_{up} \\
                \psi^\kappa_{down}
        \end{array}\right) \,\,\, ,
\label{eq4}
\ee
\ba
\left[{\bd \nabla}^2+\frac{1}{A_+}\frac{\partial A_+}{\partial r}
\left(\frac{{\bd \sigma}\cdot{\bd \ell}}{r}-\frac{\partial}{\partial r}\right)
+A_+A_-\right]\psi^\kappa_{up}({\bd r})&=&0 \,\,\, ,
\label{eq5} \\
\left[{\bd \nabla}^2+\frac{1}{A_-}\frac{\partial A_-}{\partial r}
\left(\frac{{\bd \sigma}\cdot{\bd \ell}}{r}-\frac{\partial}{\partial r}\right)
+A_+A_-\right]\psi^\kappa_{down}({\bd r})&=&0 \,\,\, ,
\label{eq6}
\ea
with
\ba
A_\pm &\equiv& \tilde{E}\pm \tilde{M} \,\,\, ,\label{eq7}\\
A_+A_- &=& \tilde{E}^2-\tilde{M}^2 = \tilde{K}^2 \,\,\, . \label{eq8}
\ea
or
\be
\tilde{K}^2=K^2+2M U_{central}
\ee
with
\be K^2=E^2-M^2\ee
and
\be
U_{central}=\frac{V^2-S^2-2EV+2MS}{2M}
\ee

Here we have defined $K^2=E^2-M^2<0$ for bound states, and a central
potential $U_{central}$ that depends on the energy $E$. Eqs.(~\ref{eq5})
and (\ref{eq6}) are Schr\"odinger--like equations for the upper and lower
components, respectively, containing in addition to the kinetic
energy and central potential ($U_{central}$) terms, a spin-orbit
($\displaystyle \frac{1}{A_{\pm}} \frac{\partial A_{\pm}}{\partial r}
\frac{{\bd \sigma}\cdot{\bd\ell}}{r}$) term and a Darwin 
($\displaystyle \frac{1}{A_{\pm}} \frac{\partial A_{\pm}}{\partial r}
\frac{\partial}{\partial r}$) term. The last two being generally
different for the upper and lower components.

We recall that in the relativistic case the bound nucleon wave function
is labeled by the quantum
number $\kappa$ that characterizes the eigenvalues of total angular
momentum and parity of the four--spinor. The relativistic parity operator 
$\Pi =\gamma^0P$, with $P\phi({\bd r})=\phi(-{\bd r})$, has the eigenvalue
$\Pi =(-1)^\kappa \frac{\kappa}{|\kappa |}$ and the $j$--eigenvalue is
$j=|\kappa|-\frac{1}{2}$. In standard notation $\psi^\kappa$ is
written as
\be
\psi^\kappa_m({\bd r})=
\left(\begin{array}{@{\hspace{0pt}}c@{\hspace{0pt}}}
                g_\kappa(r) \phi^\kappa_m(\hat{r}) \\
                if_\kappa(r)\phi^{-\kappa}_m(\hat{r})
        \end{array}\right) \label{eq9}\,\,\, ,
\ee
with
\ba
\phi^\kappa_m &=& \sum_{\Lambda \sigma}\langle \ell \Lambda \frac{1}{2} \sigma|
j m\rangle Y^\ell_\Lambda\chi_\sigma \,\,\, ,
\label{eq10} \\
\phi^{-\kappa}_m &=& -{\bd \sigma}\cdot\hat{\bd r}\phi_m^\kappa=
\sum_{\Lambda \sigma}\langle \overline{\ell} \Lambda \frac{1}{2} \sigma|
j m\rangle Y^{\overline{\ell}}_\Lambda\chi_\sigma \,\,\, .
\label{eq11}
\ea
The relationship between $\kappa$, $j$, $\ell$ and $\overline{\ell}$ is:\\
For $\kappa >0$:
\be
\kappa=j+\frac{1}{2}=\ell=\overline{\ell}+1\,\,\, , \,\,\,\,\, \Pi
=(-1)^\ell \,\,\, .\label{eq12}
\ee
For $\kappa <0$:
\be
\kappa=-\left(j+\frac{1}{2}\right)=-(\ell+1)=-\overline{\ell}\,\,\, , 
\,\,\,\,\, \Pi=(-1)^\ell \,\,\, . \label{eq13}
\ee

Because in studies of nuclear structure we are more familiar with 
the non--relativistic labeling of single--particle states, in practice 
it is convenient to resort to the $\ell_j$ characterization of the 
wave functions as in the non--relativistic limit,
where $\ell$ is a good quantum number and the spin--orbit partners are the
$j=\ell\pm\frac{1}{2}$ states. To fix ideas and to avoid confusion, in what
follows we refer to the ``stretched'' ($j=\ell+\frac{1}{2}$) 
and ``jack--knifed''
($j=\ell-\frac{1}{2}$) partners as the
states with $\kappa=-(\ell+1)<0$, and $\kappa=\ell>0$, respectively, with
$\ell$ characterizing the orbital angular momentum of the upper component
as in eqs.~(\ref{eq9}--\ref{eq13}). Hence the lower components of 
the spin--orbit partners have
$\overline{\ell}=\ell+1$ and $\overline{\ell}=\ell-1$ for the stretched
and jack--knifed $\ell_j$ partners, respectively.

The general formulas above may be cast in various forms --- here we
write them as equations for upper and lower components with effective
central and spin--orbit potentials to bring out the essential differences
to be found for the radial wave functions that result. Alternative
schemes may be used (for instance, through the introduction of an
effective mass; see, for example, \cite{Ser86}), although the present one 
is sufficient for our purposes. The Schr\"odinger--like 
equations for the upper components of the spin--orbit partners 
are:

For the stretched case ($j=\ell+1/2$),
\be
\left[\frac{1}{r}\frac{\partial^2}{\partial r^2}r -
\frac{\ell(\ell+1)}{r^2} 
+2M  U_{central} -\frac{1}{A_+}\frac{\partial A_+}{\partial r}
\left(\frac{\partial}{\partial r}-
\frac{\ell}{r}\right)\right]g_\kappa
=-K^2g_\kappa (r) \,\,\, ,
\label{eq14}
\ee
For the jack-knifed case ($j=\ell-1/2$),
\be
\left[\frac{1}{r}\frac{\partial^2}{\partial r^2}r -
\frac{\ell(\ell+1)}{r^2} 
+2M  U_{central} -\frac{1}{A_+}\frac{\partial A_+}{\partial r}
\left(\frac{\partial}{\partial r}+
\frac{\ell+1}{r}\right)\right]g_\kappa
=-K^2g_\kappa (r) \,\,\, ,
\label{eq15}
\ee
Here, in close analogy with the usual non--relativistic
limit, the equations for the spin-orbit partners differ only in their
spin-orbit coupling terms, and these are relatively small.

On the other hand the 
Schr\"odinger--like equation for the lower component
\be
\left[\frac{1}{r}\frac{\partial^2}{\partial r^2}r -
\frac{\overline{\ell}(\overline{\ell}+1)}{r^2}+2M U_{central}-
\frac{1}{A_-}\frac{\partial A_-}{\partial r}
\left(\frac{\partial}{\partial r}-
\frac{j(j+1)-\overline{\ell}(\overline{\ell}+1)-\frac{3}{4}}{r}
\right)\right]f_\kappa
=-K^2f_\kappa \,\,\, ,
\label{eq16}
\ee
when applied to the spin--orbit partners reads:

For the stretched case ($j=\ell+1/2$),
\be
\left[\frac{1}{r}\frac{\partial^2}{\partial r^2}r -
\frac{(\ell+1)(\ell +2)}{r^2}
+2M U_{central}- \frac{1}{A_-}\frac{\partial A_-}{\partial r}
\left(\frac{\partial}{\partial r}+
\frac{\ell+2}{r}\right)
\right]f_\kappa
=-K^2 f_\kappa \,\,\, ,
\label{eq17}
\ee
for the jack-knifed case ($j=\ell-1/2$)
\be
\left[\frac{1}{r}\frac{\partial^2}{\partial r^2}r -
\frac{\ell(\ell -1)}{r^2}
+2M U_{central}- \frac{1}{A_-}\frac{\partial A_-}{\partial r}
\left(\frac{\partial}{\partial r}-
\frac{\ell-1}{r}\right)
\right]f_\kappa
=-K^2 f_\kappa \,\,\, ,
\label{eq18}
\ee

Hence, in contrast to the case of the upper components, the equations for the
lower components differ not only in the spin--orbit coupling term but
also in the centrifugal barrier. This produces stronger differences on the
radial dependence of the lower components than on the
radial dependence of the upper components. This is to say, if one compares
for instance the radial functions $f_\kappa$ for the
$p_{1/2}$ and $p_{3/2}$ cases one expects to see larger differences than
if one compares the $g_\kappa$ for those same shells.

Later we shall see the consequences of these differences in upper
and lower components when examining the RPWIA results in detail; here it
should be noted that in general the lower components are small and the radial
dependence of both scalar and baryonic densities is mostly 
dominated by the upper components:
\ba
\rho_S (r) &=& \frac{4\pi}{2j+1}
 \sum_m\overline{\psi}^\kappa_m\psi^\kappa_m
= \left[g_\kappa^2(r)-f_\kappa^2(r)\right] \,\,\, ,
\label{eq19}
\\
\rho_B(r) &= & \frac{4\pi}{2j+1}
\sum_m\psi^{\kappa +}_m
\psi^\kappa_m
=\left[g_\kappa^2(r)+f_\kappa^2(r)\right] \,\,\, ,
\label{eq20}
\ea
with
\be
\int \rho_B(r)r^2dr =1\,\,\,  .
\label{eq21}
\ee

In the cases considered here the contribution of $f_\kappa$ to the
normalization is less than $3\%$. Nevertheless it is important to realize
that for spin--orbit partners the lower components can differ significantly,
while the differences in the upper components tend to be mild. This is
illustrated in fig.~1 where we plot the sum and difference of baryonic
and scalar densities (in fm$^{-3}$) 
for the spin--orbit partners $p_{1/2}$, $p_{3/2}$ in 
$^{16}$O and $d_{3/2}$, $d_{5/2}$ in $^{40}$Ca. The $3s_{1/2}$ orbital in
$^{208}$Pb is also plotted for comparison. 
The present calculations follow similar lines to those of
ref.~\cite{Cab98}; in particular we use the parameters of
ref.~\cite{Hor81} and the TIMORA code~\cite{Hor91}.
Clearly the differences of $f_\kappa$ due to the different centrifugal
barriers are maximal in the $p_{1/2}$--$p_{3/2}$ spin--orbit partners
because the lower component for the $p_{1/2}$ partner is an s--wave while
for the $p_{3/2}$ is a d--wave. The $d_{3/2}$--$d_{5/2}$ spin--orbit
partners have correspondingly a p--wave and an f--wave as lower
components. It is also interesting to see that for the $3s_{1/2}$ case the
lower component looks negligibly small and may be expected to play a minor
role. 

We note that the elementary three--vector current density
\be
{\bd v}=\overline{\psi}^\kappa {\bd \gamma}\psi^\kappa \equiv
v(r)\hat{\bd v}
\label{eq23}
\ee
is proportional to the product of $g_\kappa$ and $f_\kappa$
\be
v(r)=2g_\kappa (r) f_\kappa(r) \,\,\, .
\label{eq24}
\ee
This quantity is also plotted in fig.~2 for the same orbitals shown in
fig.~1. 

The larger differences seen for the lower components reflect a genuine
relativistic effect that cannot be accounted for when non--relativistic
bound nucleon wave functions are used. This naturally motivates an 
investigation as to whether or not this relativistic effect
has observable consequences. Electron scattering at
intermediate energies, involving large momentum transfers, is the ideal
place to look for such relativistic effects.

In elastic electron scattering one is sensitive to the Fourier
transforms of $\rho_B({\bd r})$ and of ${\bd v}({\bd r})$ and relativistic
calculations have been presented for a few elastic magnetic form
factors~\cite{magne}. A thorough discussion of relativistic
calculations of transverse form factors for elastic electron scattering
could provide the motivation for future work --- however, following our
recent studies~\cite{Cab98} 
the focus of this paper is instead placed on the longitudinal--transverse 
response functions ($R^{TL}$) for exclusive quasielastic electron 
scattering (e,e$'$p). This particular response function depends
on linear combinations of scalar, baryonic and vector densities in
momentum space
\ba
\rho_S(p)&=&\frac{4\pi}{2j+1}\sum_m\overline{\psi}^\kappa_m({\bd p})
\psi^\kappa_m({\bd p})=g_\kappa^2(p)-f_\kappa^2(p) \,\,\, ,
\label{eq25}\\
\rho_B(p)&=&\frac{4\pi}{2j+1}\sum_m\psi^{\kappa +}_m({\bd p})
\psi^\kappa_m({\bd p})=g_\kappa^2(p)+f_\kappa^2(p) \,\,\, ,
\label{eq26}\\
v(p)&=&\frac{4\pi}{2j+1}\sum_m\overline{\psi}^\kappa_m({\bd p})
\hat{\bd p}\cdot{\bd \gamma}
\psi^\kappa_m({\bd p})=2\frac{\kappa}{|\kappa|}g_\kappa(p)f_\kappa(p)
\,\,\, ,
\label{eq27}
\ea
with $\psi^\kappa_m({\bd p})$ the Fourier transform of 
$\psi^\kappa_m({\bd r})$. The radial functions in momentum space are
given by (see, for instance, \cite{Cab98})
\ba
g_\kappa (p) &=& \sqrt{\frac{2}{\pi}}\int_0^\infty r^2 dr g_\kappa (r)
j_\ell (pr) \,\,\, ,
\label{eq28}
\\
f_\kappa (p) &=& \sqrt{\frac{2}{\pi}}\int_0^\infty r^2 dr f_\kappa (r)
j_{\overline{\ell}}(pr) \,\,\, .
\label{eq29}
\ea

Specifically, the relevant quantities in p--space that enter directly
in the calculations of response functions for (e,e$'$p) processes are the
amplitudes $\tilde{\alpha}_\kappa$ and $\tilde{\beta}_\kappa$ of the
positive and negative energy projections of the bound nucleon wave
function in the expansion:
\be
\psi^\kappa_m({\bd p})=\sum_s\left[\tilde{\alpha}_\kappa (p)
u(p,s)\langle s|i^\ell \phi^\kappa_m(\hat{\bd p})\rangle+
\tilde{\beta}_\kappa (p)
v(p,s)\langle s|i^\ell \phi^{-\kappa}_m(\hat{\bd p})\rangle
\right] \,\,\, ,
\label{eq30}
\ee
with
\ba
\tilde{\alpha}_\kappa (p)&=&\sqrt{\frac{\overline{E}+M}{2M}}
\left(g_\kappa(p)-\frac{\kappa}{|\kappa|}f_\kappa(p)\frac{p}{\overline{E}+M}
\right)\,\,\, , \label{eq31} \\
\tilde{\beta}_\kappa (p)&=&\sqrt{\frac{\overline{E}+M}{2M}}
\left(\frac{p}{\overline{E}+M}g_\kappa(p)-\frac{\kappa}{|\kappa|}f_\kappa(p)
\right) \,\,\, , \label{eq32}
\ea
where $u(p,s)$ and $v(p,s)$ are the positive and negative energy solutions
of the free Dirac equation with $\overline{E}=\sqrt{M^2+p^2}$. The
absolute values of the positive and negative energy 
projection amplitudes (in fm$^3$)
are plotted in fig.~3 for the spin--orbit partners $p_{1/2}$--$p_{3/2}$ and
$d_{3/2}$--$d_{5/2}$. 
Note the different scales of the positive and
negative energy projections. Note also that in the limit of free nucleons
$\tilde{\beta}_\kappa (p)=0$, and therefore $|\tilde{\beta}_\kappa |^2$ 
measures the dynamical enhancement of the lower components.

A much larger dynamical enhancement is observed in fig.~3 for the
jack-knifed states than for the stretched states. This was expected from
our previous discusion on the Schr\"odinger--like equations for the upper
and lower components.
As a matter of fact, much can be learnt by looking at the
Schr\"odinger--like equations that follow from the Dirac eq.~(\ref{eq1}).
For instance, as first pointed ouy by Ginocchio~\cite{Gin97}
the Schr\"odinger--like equation for the lower component
is the same for relativistic states having the same
$\overline{\ell}$--value but $\ell$--values differing by two units (as
for instance the $s_{1/2}$--$d_{3/2}$ or $p_{3/2}$--$f_{5/2}$,...
pairs) when the spin--orbit coupling term in eq.~(\ref{eq16}) is
negligible. This has been identified as the origin of pseudospin symmetry
in nuclear spectra by Ginocchio~\cite{Gin97}. Another interesting
limiting case can be that of quasidegeneracy of levels with equal
$j$ and oposite parity that is also present in nuclear spectra~\cite{Lede}.

\section*{3. Longitudinal--transverse response function}

In this section we compare results on the $R^{TL}$ response functions for
one--nucleon knockout from various relativistic bound orbitals. The
details of the calculations are as in ref.~\cite{Cab98} and we do not
repeat them here; we use what was called kinematics I in that work, 
namely quasielastic conditions with $q=$ 500 MeV/c and $\omega=$ 131.56
MeV. We recall that, using projection techniques, the $R^{TL}$
response can be separated into a contribution from the positive energy
projection ($R^{TL}_P$), a contribution from the negative energy projection
($R^{TL}_N$), and a crossed term ($R^{TL}_C$):
\be
R^{TL}=R^{TL}_P+R^{TL}_N+R^{TL}_C \,\,\, ,
\label{eq33}
\ee
with
\ba
R^{TL}_P&=& {\cal R}^{TL}_{uu}N_{uu}(p) \,\,\, , \label{eq34} \\
R^{TL}_N&=& {\cal R}^{TL}_{vv}N_{vv}(p) \,\,\, , \label{eq35} \\
R^{TL}_C&=& {\cal R}^{TL}_{uv}N_{uv}(p) \,\,\, . \label{eq36} 
\ea
In eqs.~(\ref{eq34}--\ref{eq36}) the dependence on the nuclear structure
is factorized. This dependence is contained in the bound
nucleon momentum distributions
\ba
N_{uu}(p)&=&\left(\tilde{\alpha}_\kappa(p)\right)^2/4\pi \,\,\, ,
\label{eq37} \\
N_{vv}(p)&=&\left(\tilde{\beta}_\kappa(p)\right)^2/4\pi \,\,\, ,
\label{eq38} \\
N_{uv}(p)&=&-2\tilde{\alpha}_\kappa(p)\tilde{\beta}_\kappa (p)/4\pi
\,\,\, . \label{eq39}
\ea
In contrast, ${\cal R}^{TL}_{uu}$, ${\cal R}^{TL}_{vv}$ and
${\cal R}^{TL}_{uv}$ depend only on the current operator and gauge and are
independent of the nuclear structure.

When the nuclear structure is treated in the non--relativistic limit
the nucleon current operator is expanded in the basis of positive
energy free Dirac spinors (see for instance ref.~\cite{Fru84}) and the
single--nucleon response function is
\be
{\cal R}^{TL}={\cal R}^{TL}_{uu} \,\,\, , \label{eq40}
\ee
corresponding to electron scattering from a free (relativistic) nucleon,
and 
\be
R^{TL} \stackrel{n.r.}{\longrightarrow} R^{TL}_P \,\,\,\, .\label{eq41}
\ee
With relativistic bound nucleon wave functions one has additional
single--nucleon responses ${\cal R}_{uv}^{TL}$ and ${\cal R}_{vv}^{TL}$
that do not appear in the scattering of electrons from free nucleons (or
antinucleons). The appearance of these terms is due to the nonzero overlap
of the bound nucleon wave function with the Dirac sea that results in
finite values of the $N_{uv}(p)$ and $N_{vv}(p)$ functions.

In ref.~\cite{Cab98} we showed that the single--nucleon responses 
${\cal R}_{uv}^{TL}$ and ${\cal R}_{vv}^{TL}$ change much more with the
changes of the current operator and gauges than does the
${\cal R}_{uu}^{TL}$ response. Whereas the single--nucleon response
${\cal R}_{uu}^{TL}$ has often been studied~\cite{Cab93}, the
${\cal R}_{uv}^{TL}$ and ${\cal R}_{vv}^{TL}$ were for the first time
identified and studied in ref.~\cite{Cab98}. The fact that these responses
are more strongly dependent on the choice of the current operator and
gauge, results in a larger theoretical ambiguity in the calculated total
response $R^{TL}$ when using relativistic bound nucleon wave functions.

As we showed in the previous section, the overlap with the Dirac sea of the
bound nucleon wave function depends strongly on the particular 
$\ell_j$ orbital
under study. Therefore a different sensitivity of the $R^{TL}$ response to
the contributions from the negative energy projections can be expected for
different orbitals. In particular we show in figs.~4 and 5 that this
sensitivity is quite different for the spin--orbit partners $p_{1/2}$ and
$p_{3/2}$ or $d_{3/2}$ and $d_{5/2}$.

The longitudinal transverse response functions for one--nucleon knockout
from the $p_{1/2}$ and $p_{3/2}$ orbitals in $^{16}$O are shown in the top
and bottom panels of fig.~4. For each shell results corresponding to the
CC1 current operator and to the three different gauges (Landau, Coulomb
and Weyl) are shown on the left panels, while the right hand panels
contain the corresponding results for the CC2 current operator.

In these figures one can see that for fixed current operator and shell the
Landau and Coulomb gauges give practically the same results, while the Weyl
gauge  tends to produce important
deviations. The importance of these deviations is much higher for
the $p_{1/2}$ than for the $p_{3/2}$ shell. This 
is also the case when changing the current operator.
 The following ratios are found
between the maximum values of the $R^{TL}$ response of the $p_{1/2}$
shell: if we fix the gauge to be the Coulomb gauge, the ratio between the
results for the CC1 and CC2 current operators is
\be
\frac{R^{TL}(CC1(0))}{R^{TL}(CC2(0))}\simeq 1.5 \,\,\, ,
\ee
and the ratio is larger in the Weyl gauge
\be
\frac{R^{TL}(CC1(3))}{R^{TL}(CC2(3))}\simeq 2 \,\,\, .
\ee
Similar ratios are found when we fix the current operator and compare the
results of different gauges
\be
\frac{R^{TL}(CC2(3))}{R^{TL}(CC2(0))}\simeq 1.5 \,\,\, ,
\,\,\,\,\,\,\,\,\,\,\,\,\,\,\,\,\,\,
\frac{R^{TL}(CC1(3))}{R^{TL}(CC1(0))}\simeq 2 \,\,\, ,
\ee
leading to the large difference, by as much as a factor of 3, between the
CC1(3) and the CC2(0) results, already mentioned in ref.~\cite{Cab98}.

On the contrary, for the spin--orbit partner $p_{3/2}$ the results between
different choices are not that large so the above ratios reach at most a
value of 1.2 (that between CC1(3) and CC2(0) results) and in most cases
are less than a $10\%$. In particular for the $p_{3/2}$ case the
dependence on the gauge is almost not visible  with the CC2
current operator.

We note that $R^{TL}$ contains the product of
transverse and longitudinal currents. The transverse current changes only
with the current operator (i.e., CC1 or CC2) but it is independent on the
gauge. Therefore, for the cases considered, the strong gauge dependence 
for the $p_{1/2}$ case must
come from the longitudinal current. A similar comparison is made in fig.~5
for the $d_{3/2}$ and $d_{5/2}$ spin--orbit partners. It is also the case
that the dependence on the current operator and gauge is more pronounced
for the jack--knifed than for the stretched cases.

As seen in figs.~6 and 7 when we consider only the contribution from the
positive energy projection (i.e., the truncation in eq.~(\ref{eq41})) the
sensitivity of the $R^{TL}$ response to the choice of the current operator
and gauge is similar for the spin--orbit partners $p_{1/2}$--$p_{3/2}$ and
$d_{3/2}$--$d_{5/2}$. The larger sensitivity obtained for the $p_{1/2}$
and $d_{3/2}$ partners is due to their larger negative energy projections
at low $p$ ($p\leq 200$ MeV) that carries larger deviations particularly in
the $R^{TL}_C$ term. This is in turn an effect of the Schr\"odinger--like
equations for the lower components of the spin--orbit partners.\\

\section*{4. Final remarks}

The main conclusion of this work is that the large deviation~\cite{Cab98}
in $R^{TL}$ predictions produced by different current operators and gauges
takes place for the jack--knifed states ($p_{1/2}$, $d_{3/2}$,...), but
not for the stretched states ($p_{3/2}$, $d_{5/2}$,...).
This genuine relativistic effect stems from the
dynamical enhancement of the lower components, and can be traced back
to the Schr\"odinger--like equations for the upper and lower components 
of the relativistic bound nucleon wave function.

We have shown that this effect  can be understood from the quite different
behaviour of the dynamical enhancement function $\tilde{\beta}_\kappa$
of the stretched and jack-knifed states. Indeed, from eq. (\ref{eq32})
and fig. 3 one can see that from the jack-knifed states ({\em i.e.}
$p_{1/2}$, $d_{3/2}$, $\dots$) the amplitudes of the negative energy
projections are much larger than those for the stretched states
($p_{3/2}$, $d_{5/2}$, $\dots$). One has to realize that in eq.
(\ref{eq32}) $\tilde{\beta}_\kappa(p)$ goes like the difference of
two functions of $p$ with the same number of nodes ($n+\ell+1$) for the
stretched states, while for the jack-knifed ones, it goes like the
difference of a function with $n+\ell+1$ nodes and a function with
$n+\ell-1$ nodes ($n=1$ in the cases considered here). This is due 
to the different quantum number $\bar{\ell}$ of the lower components in
stretched and jack-knifed states and can be traced back to the
Schr\"odinger--like equations (\ref{eq17})--(\ref{eq18}). This is in
contrast to the amplitudes of the positive energy projections
($\tilde{\alpha}_\kappa$) that in all cases are strongly dominated by the
upper component ($g_\kappa$) and are therefore similar for stretched and
 jack-knifed states. Note that $g_\kappa(r)$ for stretched and jack-knifed
states (with equal $n$ and $\ell$) differ only in their {\em r.m.s.} radii.
 
Since the non--relativistic limit corresponds to
$\tilde{\beta}_\kappa(p)=0$, as a corolary we may also conclude that
comparing to the non--relativistic limit, the responses for the stretched
states will be closer  than those for the jack-knifed states to their
respective non--relativistic limits.

The choice of current operator and gauge is an important issue in
quasielastic electron scattering since theoretical results on differential
cross-sections and response functions depend substantially on those
choices (see for instance refs.~\cite{Naus,Fru84,Cab93} and refs.
therein). This tends to be particularly so when relativistic bound and/or
scattering wave functions are used because the contributions from the
negative energy projections depend more strongly on the current operator
choice than do the contributions from the positive energy
projections. The relativistic approach provides a tool to
study further reaching consequences of the various choices of the current
operators and gauges that can not be explored if one attaches to the 
 non--relativistic description
of the nuclear structure and wave functions where one can not go
beyond the truncated expression in eq.~(\ref{eq41}).

In the case of stretched states the differences in the $R^{TL}$
responses with different current operators and gauges are of the same
order (and have the same origin) as those found with non--relativistic bound
nucleon wave functions. The theoretical uncertainty for these states is at
most a $20\%$ (this is the largest ratio between CC1(3) and CC2(0)
calculations), which is small compared to the large theoretical
uncertainty of up to $300\%$ found for the jack--knifed states.
Although our calculations here do not contain final--state interactions, a
similar behaviour is to be expected when FSI are included. Since FSI and
short--range initial--state correlations are expected to play a similar role 
when calculating the $R^{TL}$ response of different spin--orbit partners, the
effect found here may provide a way to test different choices of current
operators and gauges. 

Since the results for jack-knifed states amplify
differences between models, while those for the stretched states are more
``model independent'', the combined analysis of data on spin-orbit
partners can be used to elucidate between models. On the other hand,
if one wants to determine with minimal model dependence a certain
observable, one should focus on measurements on stretched states.



\subsection*{Acknowledgements}
This work is supported in part by a NATO Collaborative Research Grant
Number 940183, in part by DGICYT (Spain) under Contract Nos. PB/95--0123
and PB/95--0533--A, in part by Complutense University (Madrid) under
Project No. PR156/97
and in part by funds provided by the US Department of
Energy (D.O.E.) under cooperative agreement \#DE--FC01--94ER40818.


\newpage
\section*{Figure captions}
\begin{enumerate}
\item[Figure 1:] Semi--sum (left panels) and semi--difference (right panels) 
of the
baryonic and scalar densities. Results are shown for the spin--orbit partners:
$1p_{1/2}$ (thick--dotted line), $1p_{3/2}$ (thin--dotted line), and
$1d_{3/2}$ (thick--dashed), $1d_{5/2}$ (thin--dashed). For comparison we also
present results for the shell $3s_{1/2}$ (solid line).

\item[Figure 2:] Same as fig.~1 for the vector current density
(eq.~(\ref{eq24})).

\item[Figure 3:] Amplitudes $|\tilde{\alpha}_\kappa (p)|$ (top panel)
and $|\tilde{\beta}_\kappa (p)|$ (bottom panel) for the shells:
$1p_{1/2}$, $1p_{3/2}$, $1d_{3/2}$ and $1d_{5/2}$.
The labeling of the various curves is the same as in fig.~1.

\item[Figure 4:] Interference longitudinal--transverse response function
$R^{TL}$ (in units of fm$^3$)
for the shells: $1p_{1/2}$ (top panels) and
$1p_{3/2}$ (bottom panels) in $^{16}$O. Panels on the left hand side
correspond to results for the CC1 current operator and the gauges:
Landau (solid line), Coulomb (dotted) and Weyl (dashed). Right hand
panels show the results for the CC2 current operator and the same labeling.

\item[Figure 5:] Same as fig.~4 for the spin--orbit partners:
$1d_{3/2}$ and $1d_{5/2}$.

\item[Figure 6:] Components of the transverse--longitudinal response,
$R^{TL}_P$, $R^{TL}_C$ and $R^{TL}_N$ (see eqs.~(\ref{eq34}--\ref{eq36})). 
Results are shown for the spin--orbit partners: $1p_{1/2}$ (top panel) and
$1p_{3/2}$ (bottom panel). Thick lines correspond to gauges that use the
CC2 current operator and thin lines correspond to the CC1 operator. Gauges
considered are: Landau (solid lines), Coulomb (dotted lines) and Weyl
(dashed lines).

\item[Figure 7:] Same as fig.~6 for the shells:
$1d_{3/2}$ (top panel) and $1d_{5/2}$ (lower panel).

\end{enumerate}

\begin{figure}[t]
\begin{center}
\mbox{\epsfig{file=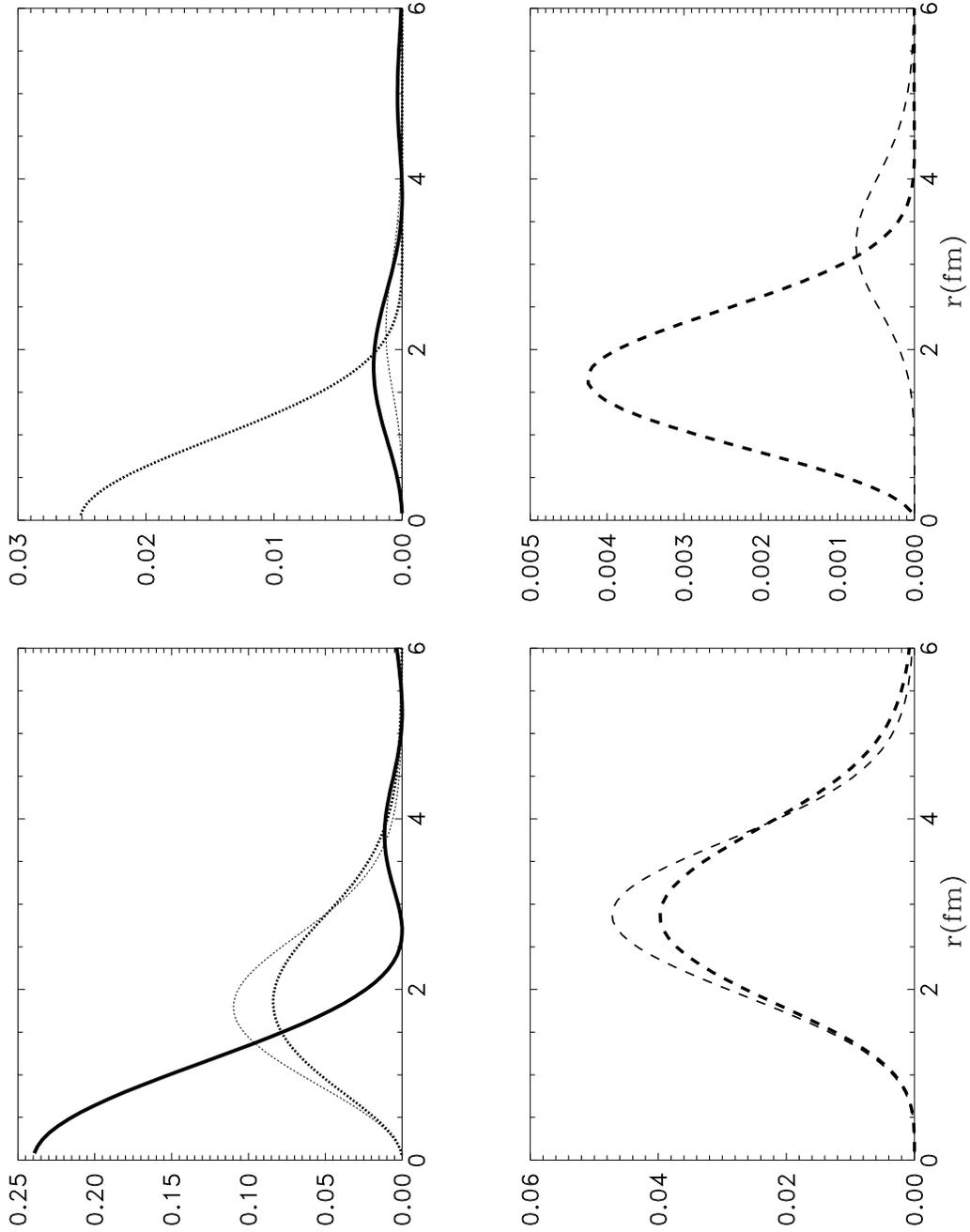,width=0.85\textwidth}}
\caption{Semi--sum (left panels) and semi--difference (right panels) of the
baryonic and scalar densities. Results are shown for the spin--orbit partners:
$1p_{1/2}$ (thick--dotted line), $1p_{3/2}$ (thin--dotted line), and
$1d_{3/2}$ (thick--dashed), $1d_{5/2}$ (thin--dashed). For comparison we also
present results for the shell $3s_{1/2}$ (solid line).}
\end{center}
\end{figure}

\begin{figure}[t]
\begin{center}
\mbox{\epsfig{file=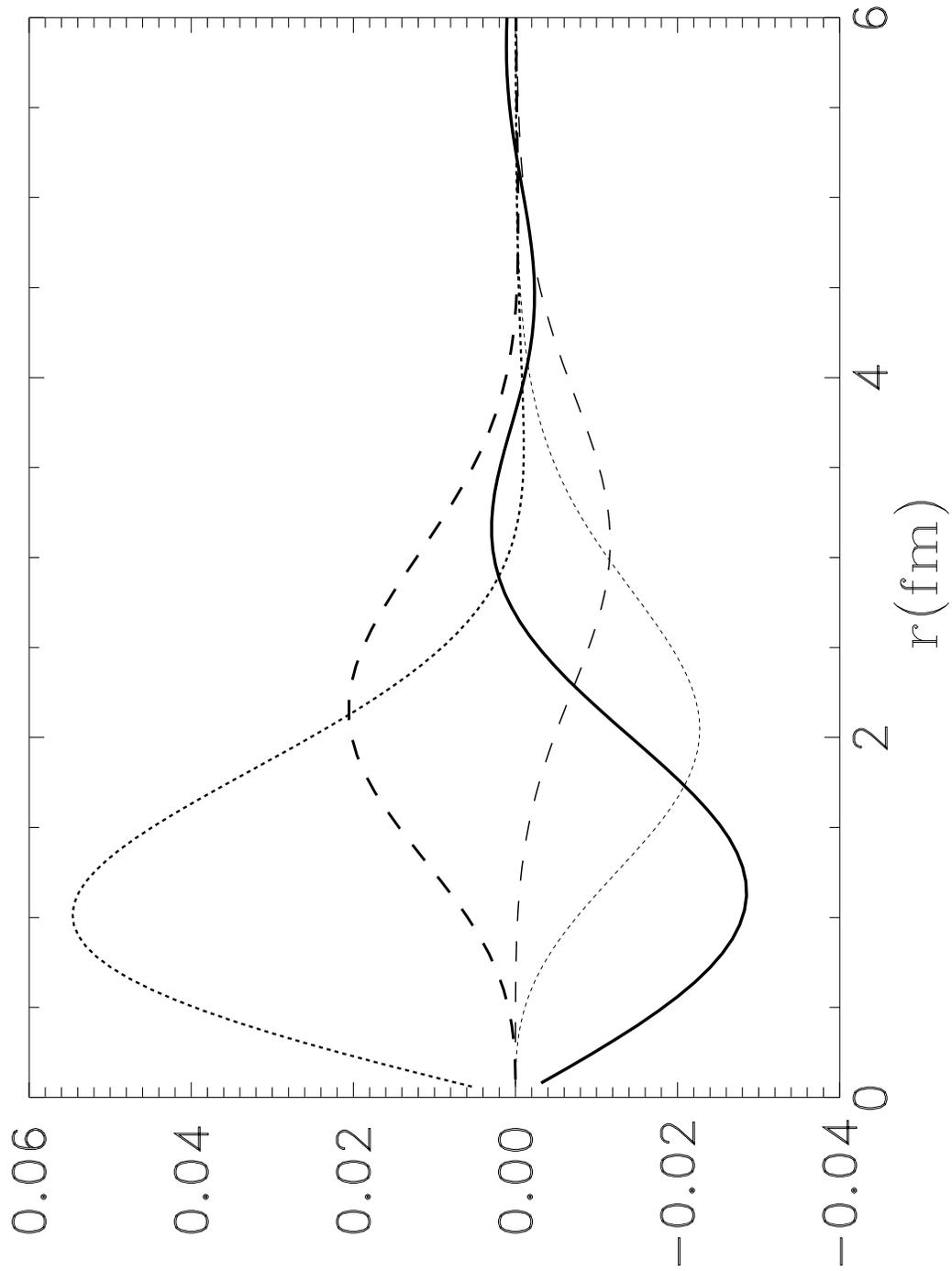,width=0.85\textwidth}}
\caption{Same as fig.~1 for the vector current density
(eq.~(\ref{eq24})).}
\end{center}
\end{figure}

\begin{figure}[t]
\begin{center}
\mbox{\epsfig{file=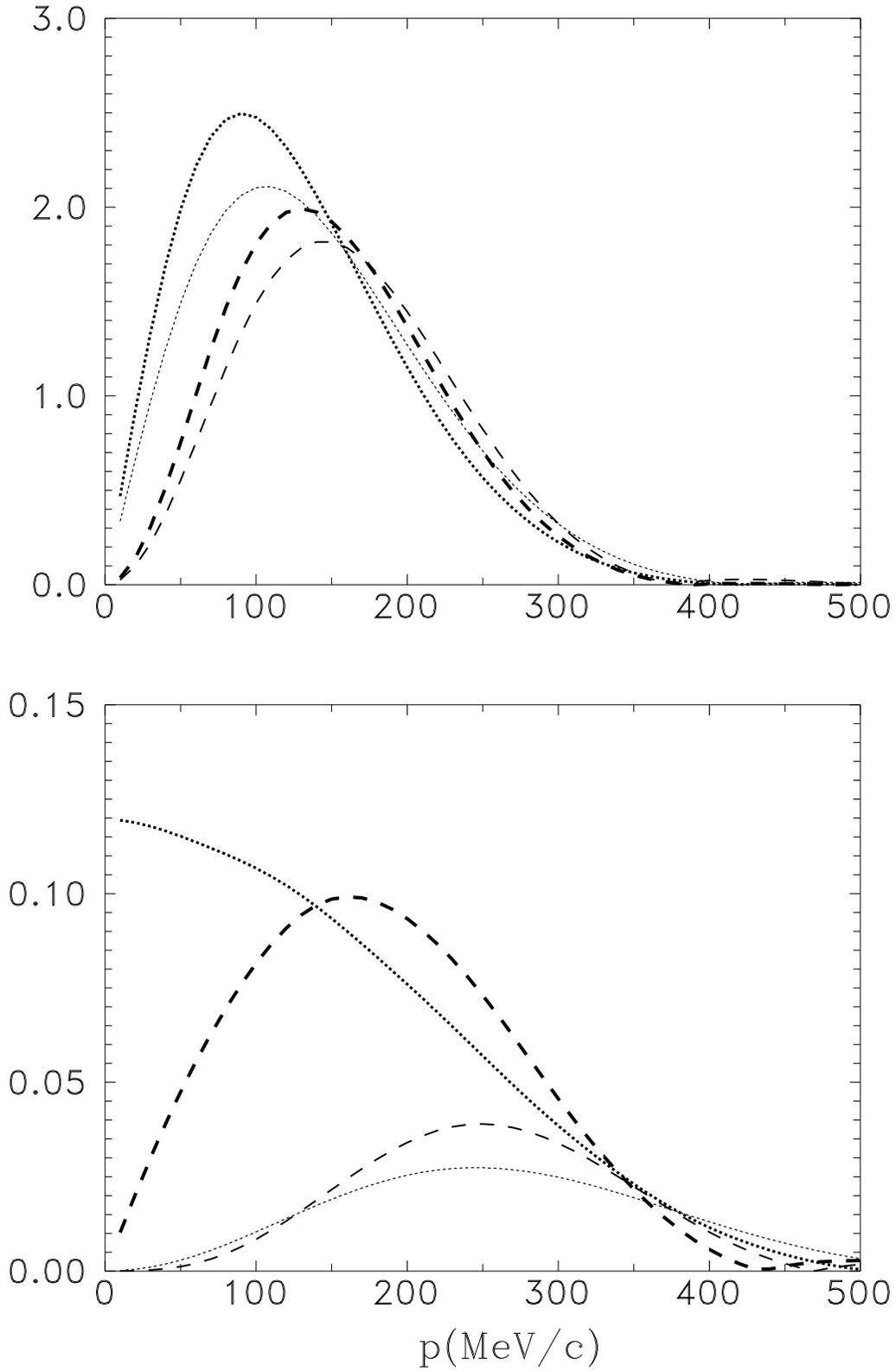,width=0.85\textwidth}}
\caption{Amplitudes $|\tilde{\alpha}_\kappa (p)|$ (top panel)
and $|\tilde{\beta}_\kappa (p)|$ (bottom panel) for the shells:
$1p_{1/2}$, $1p_{3/2}$, $1d_{3/2}$ and $1d_{5/2}$.
The labeling of the various curves is the same as in fig.~1.}
\end{center}
\end{figure}

\begin{figure}[t]
\begin{center}
\mbox{\epsfig{file=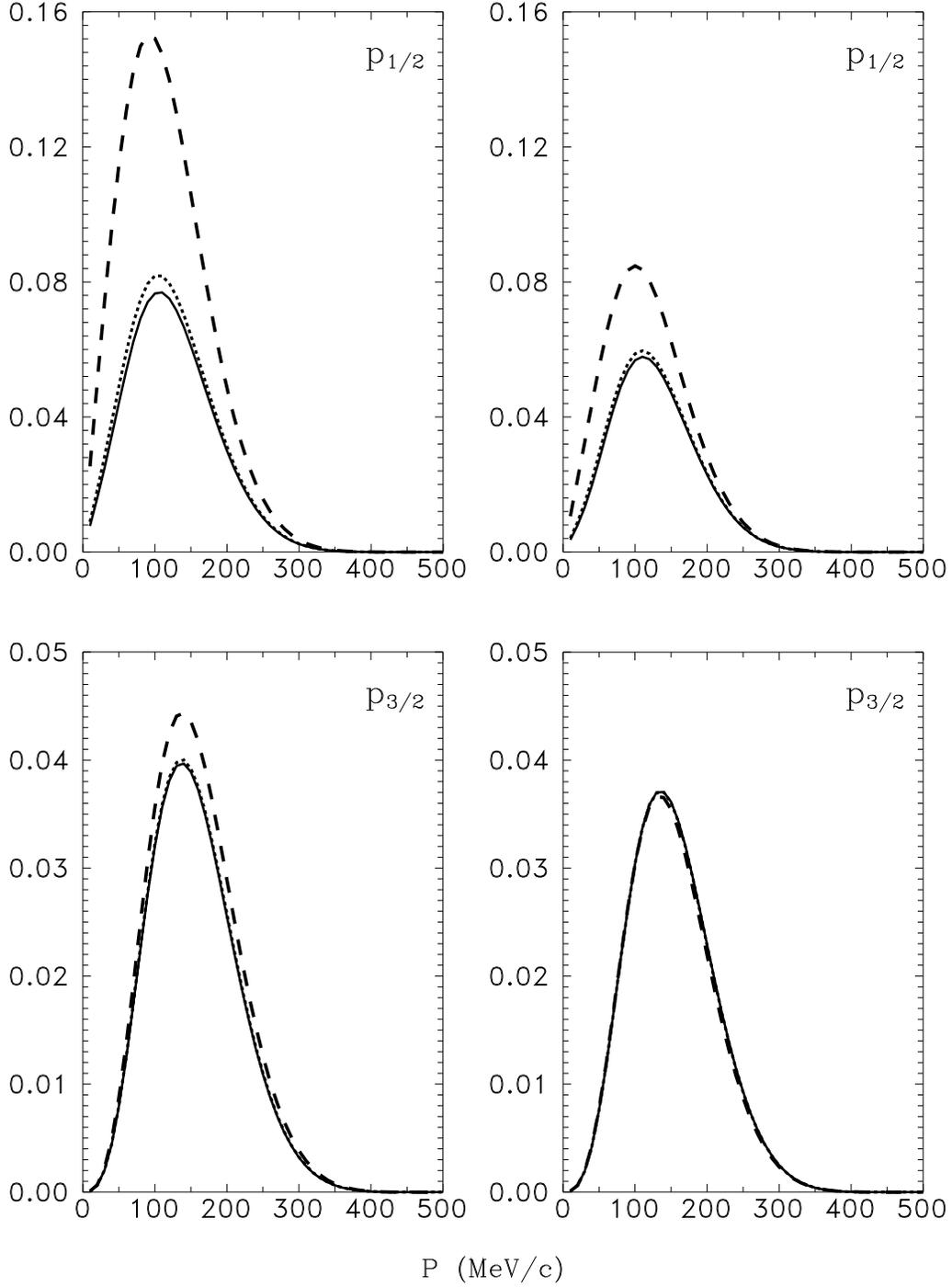,width=0.85\textwidth}}
\caption{Interference longitudinal--transverse response function
$R^{TL}$ (in units of fm$^3$)
for the shells $1p_{1/2}$ (top panels) and
$1p_{3/2}$ (bottom panels) in $^{16}$O. Panels on the left hand side
correspond to results for the CC1 current operator and the gauges:
Landau (solid line), Coulomb (dotted) and Weyl (dashed). Right hand
panels show the results for the CC2 current operator and the same labeling.}
\end{center}
\end{figure}

\begin{figure}[t]
\begin{center}
\mbox{\epsfig{file=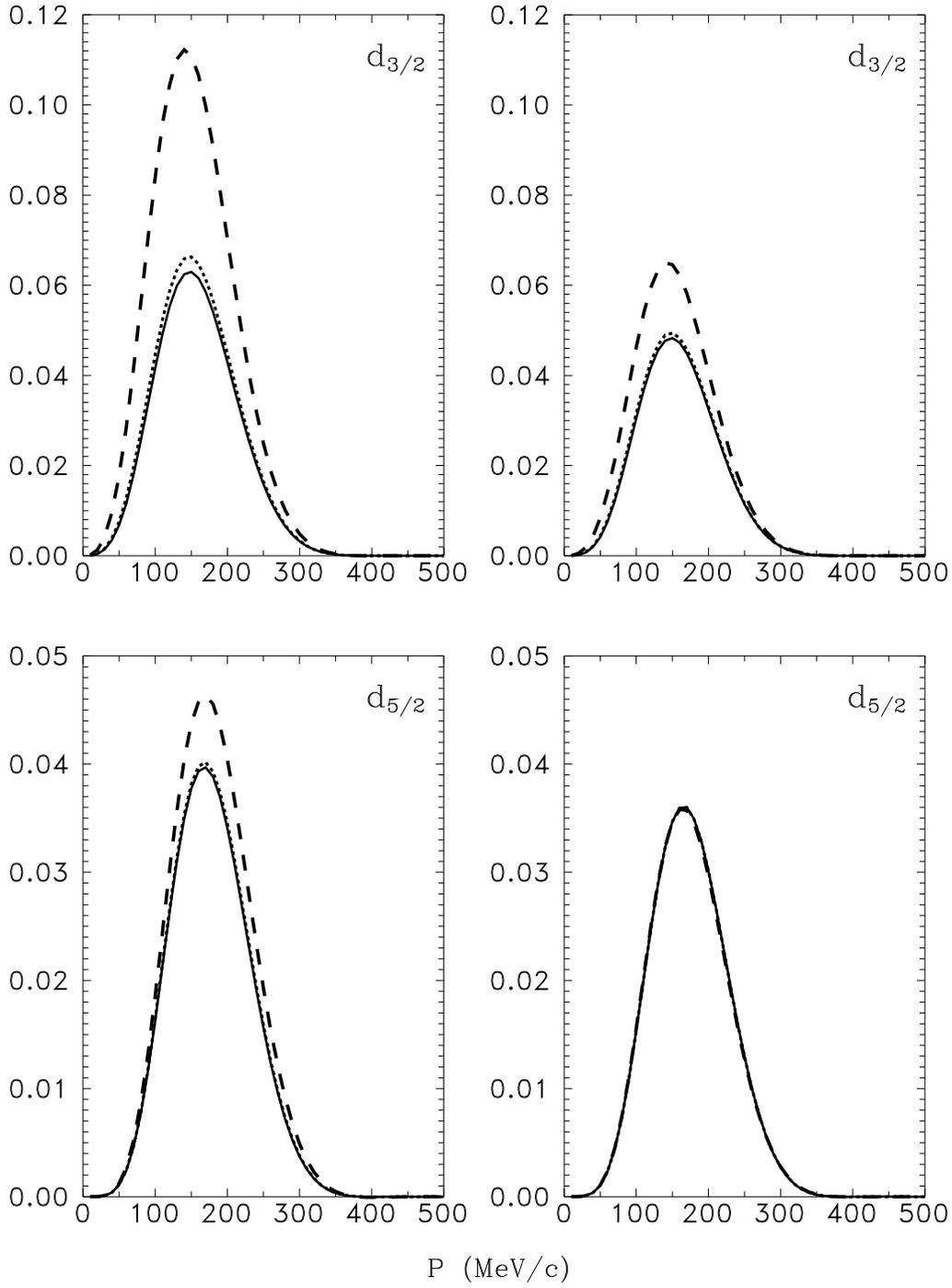,width=0.85\textwidth}}
\caption{Same as fig.~4 for the spin--orbit partners:
$1d_{3/2}$ and $1d_{5/2}$.}
\end{center}
\end{figure}

\begin{figure}[t]
\begin{center}
\mbox{\epsfig{file=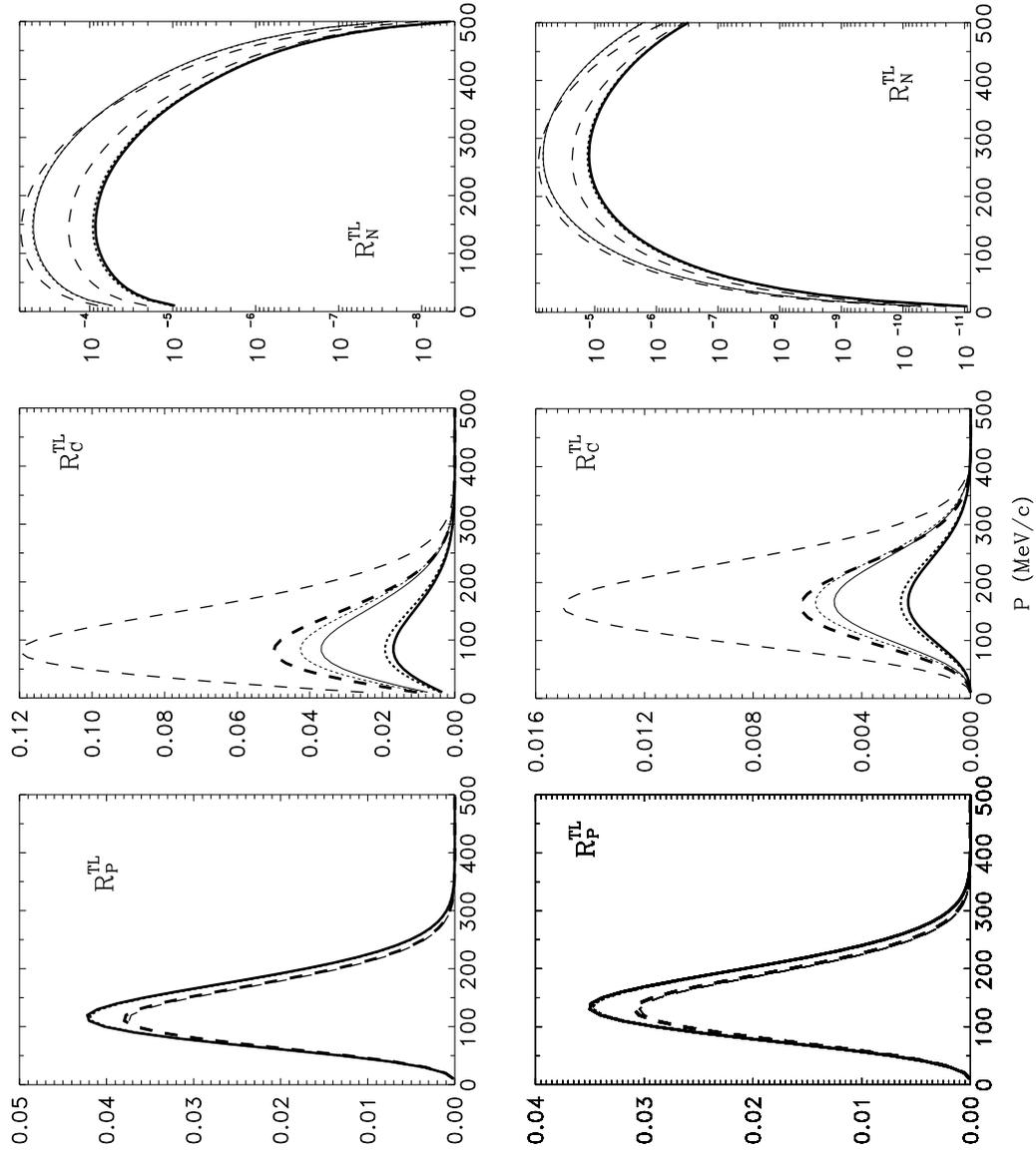,width=0.85\textwidth}}
\caption{Components of the transverse--longitudinal response,
$R^{TL}_P$, $R^{TL}_C$ and $R^{TL}_N$ (see eqs.~(\ref{eq34}--\ref{eq36})). 
Results are shown for the spin--orbit partners: $1p_{1/2}$ (top panel) and
$1p_{3/2}$ (bottom panel). Thick lines correspond to gauges that use the
CC2 current operator and thin lines correspond to the CC1 operator. Gauges
considered are: Landau (solid lines), Coulomb (dotted lines) and Weyl
(dashed lines).}
\end{center}
\end{figure}

\begin{figure}[t]
\begin{center}
\mbox{\epsfig{file=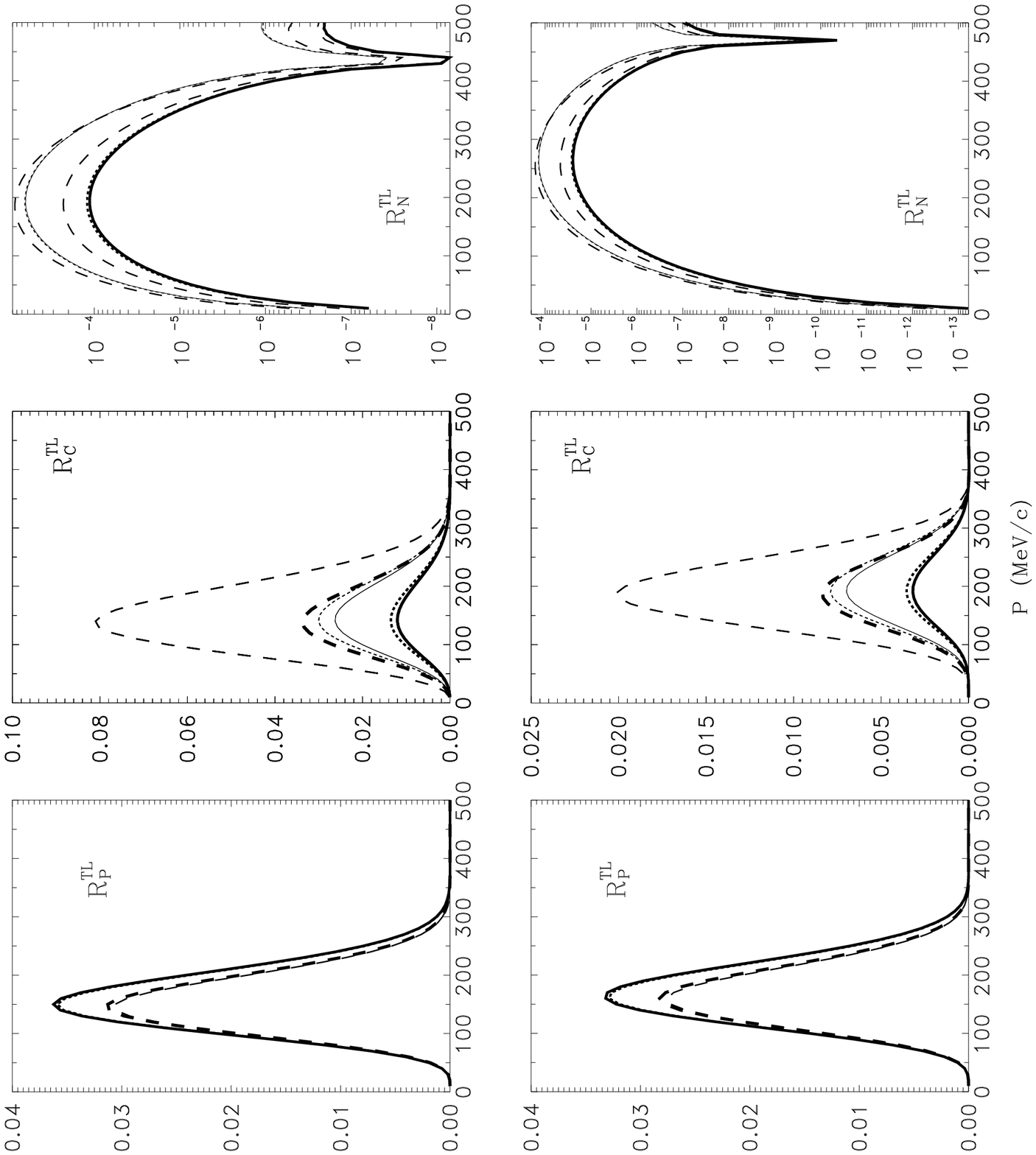,width=0.85\textwidth}}
\caption{Same as fig.~6 for the shells:
$1d_{3/2}$ (top panel) and $1d_{5/2}$ (lower panel).}
\end{center}
\end{figure}


\begin{thebibliography}{Expo92}




\bibitem{Cab98} J.A. Caballero, T.W. Donnelly, E. Moya de Guerra and J.M.
                Ud\'{\i}as,
                {\sl Nucl. Phys. A632} (1998) 323.

\bibitem{data}  I. Bobeldijk et al., {\sl Phys. Rev. Lett 73} (1994) 2684;
                J. Wesseling et al., {\sl Phys. Rev. C55} (1997) 2773;
                L. Lapik\'as, {\sl Nucl. Phys. A553} (1993) 297C.

\bibitem{Udi93} J.M. Ud\'{\i}as, P. Sarriguren, E. Moya de Guerra, E.
                Garrido and J.A. Caballero,
                {\sl Phys. Rev. C48} (1993) 2731.

\bibitem{Udi95} J.M. Ud\'{\i}as, P. Sarriguren, E. Moya de Guerra, E.
                Garrido and J.A. Caballero,
                {\sl Phys. Rev. C51} (1995) 3246.

\bibitem{Jin92} Y. Jin, D.S. Onley and L.E. Wright,
                {\sl Phys. Rev. C45} (1992) 1311.

\bibitem{Wal74} J.D. Walecka, {\sl Ann. Phys. (N.Y.) 83} (1974) 491.

\bibitem{Ser86} B.D. Serot and J.D. Walecka, 
                {\sl Adv. Nucl. Phys. 16} (1986) 1.


\bibitem{Udi96} J.M. Ud\'{\i}as, P. Sarriguren, E. Moya de Guerra and
                J.A. Caballero,
                {\sl Phys. Rev. C53} (1996) R1488.

\bibitem{npa15} S. Gardner and J. Piekarewicz, {\sl Phys. Rev. C50}
(1994) 2822.

\bibitem{npa16} A. Picklesimer, J.W. Van Orden and S.J. Wallace, {\sl
Phys.  Rev. C32} (1985) 1312;
                A. Picklesimer and J.W. Van Orden, {\sl Phys. Rev. C35}
(1987) 266; {C40} (1989) 290.

\bibitem{npa14} M. Hedayati-Poor, J.I. Johansson and H.S. Sherif, {\sl Phys 
Rev C51} (1995) 2044.

\bibitem{deF83} T. de Forest,
                {\sl Nucl. Phys. A392} (1983) 232.

\bibitem{Naus} H.W.L. Naus, S.J. Pollock, J.H. Koch and U. Oelfke,
                {\sl Nucl. Phys. A509} (1990) 717;
                S. Pollock, H.W.L. Naus and J.H. Koch,
                {\sl Phys. Rev. C53} (1996) 2304.

\bibitem{TL1}   H.J. Bulten, Ph.D. Thesis, University of Utrecht
                (1992); L. Lapik\'as, {\sl Nucl. Phys. A553} (1993) 297c;
                G.M. Spaltro et al., {\sl Phys. Rev. C48} (1993) 2385.

\bibitem{TL2}   L. Chinitz et al., {\sl Phys. Rev. Lett. 67} (1991) 568.


\bibitem{progress} 
 J.A. Caballero, T..W. Donnelly, E. Moya de Guerra and J.M.
                Ud\'{\i}as, {\sl work in progress}.
\bibitem{BD64}  J.D. Bjorken and S.D. Drell,
                in   {\sl Relativistic Quantum Mechanics}, 
             McGraw-Hill (1964).

\bibitem{Hor81} C.J. Horowitz and B.D. Serot,
                {\sl Nucl. Phys. A368} (1981) 503;
                {\sl Phys. Lett. B86} (1979) 146.

\bibitem{Hor91} C.J. Horowitz, D.P. Murdock, and B.D. Serot, in
                {\sl Computational Nuclear Physics,}
                Springer--Verlag, Berlin, 1991.

\bibitem{magne} E.J. Kim, {\sl Phys. Lett. B174} (1986) 233;
                A.O. Gattone and J.P. Vary, {\sl Phys. Lett. B219} (1989) 22.

\bibitem{Gin97} J.N. Ginocchio, {\sl Phys. Rev. Lett. 78}, (1997) 436;
                J.N. Ginocchio and D.G. Madland, {\sl Phys. Rev. C57}
                (1998) 1167.

\bibitem{Lede}  C.M. Lederer and V.S. Shirley (Eds.), in {\sl Table of
isotopes}, John Wiley and Sons, New York (1978).

\bibitem{Fru84} S. Boffi, C. Giusti and F. D. Pacati, 
                Phys. Rep. 226 (1993), 1;
                S.Frullani and J. Mougey, in 
                {\sl Advances in Nuclear  Physics},  
                edited by J.W. Negele and E. W. Vogt, 
                volume 14, Plenum Press, New York (1984).

\bibitem{Cab93} J.A. Caballero, T.W. Donnelly and G.I. Poulis,
                {\sl Nucl. Phys. A555} (1993) 709;
                J.A. Caballero, T.W. Donnelly, G.I. Poulis, E. Garrido and
                E. Moya de Guerra, {\sl Nucl. Phys. A577} (1994) 528;
                E. Garrido, J.A. Caballero, E. Moya de Guerra, P.
                Sarriguren and J.M. Ud\'{\i}as,
                {\sl Nucl. Phys. A584} (1995) 256.





































                







\end{thebibliography}
\end{document}